\newif\ifLINUXBUILD%
\IfFileExists{/dev/null}{\LINUXBUILDtrue}{}

\documentclass[fleqn,10pt]{wlscirep}
\usepackage[utf8]{inputenc}

\providecommand\citep{}
\renewcommand{\citep}[1]{\cite{#1}}
\providecommand\citet{}
\renewcommand{\citet}[1]{\cite{#1}}

\author[1,*]{Juha Nurmi}
\author[1]{Arttu Paju}
\author[2]{Billy Bob Brumley}
\author[3]{Tegan Insoll}
\author[3]{Anna K. Ovaska}
\author[3]{Valeriia Soloveva}
\author[3]{Nina Vaaranen-Valkonen}
\author[4]{Mikko Aaltonen}
\author[5]{David Arroyo}

\affil[1]{Tampere University, Tampere, FI-33720, Finland}
\affil[2]{Rochester Institute of Technology, Rochester, NY, 14623-5608, USA} %
\affil[3]{Suojellaan Lapsia, Protect Children ry., Helsinki, FI-00580, Finland}
\affil[4]{University of Eastern Finland, Joensuu, FI-80101, Finland}
\affil[5]{Consejo Superior de Investigaciones Cient\'{\i}ficas, Madrid, 28014, Spain}

\affil[*]{juha.nurmi@tuni.fi}

\usepackage[T1]{fontenc}
\usepackage{graphicx}
\usepackage{xspace}
\usepackage{lipsum}
\usepackage{booktabs}
\usepackage{multirow}

\newcounter{rqcounter}

\usepackage{siunitx}
\newcommand{\rounding}[1]{%
  \num[round-mode=places, round-precision=1]{#1}%
}

\makeatletter
\newcommand{\rqcounterautorefname}{RQ\@gobble}
\makeatother

\newcommand{\AHMIA}{\href{https://ahmia.fi/}{Ahmia.fi}}

\newcommand{\TITLE}{Investigating child sexual abuse material availability, searches, and users on the anonymous Tor network for a public health intervention strategy}

\usepackage{xcolor}

\title{\TITLE{}}

\begin{abstract}
Tor is widely used for staying anonymous online and accessing onion websites;
unfortunately, Tor is popular for distributing and viewing illicit child sexual abuse material (CSAM).
From 2018 to 2023, we analyse 176,683 onion domains and find that one-fifth share CSAM\@.
We find that CSAM is easily available using 21 out of the 26 most-used Tor search engines.
We analyse 110,133,715 search sessions from the \AHMIA{} search engine and discover that 11.1\% seek CSAM\@.
When searching CSAM by age, 40.5\% search for 11-year-olds and younger; 11.0\% for 12-year-olds;
8.2\% for 13-year-olds; 11.6\% for 14-year-olds; 10.9\% for 15-year-olds; and 12.7\% for 16-year-olds.
We demonstrate accurate filtering for search engines, introduce intervention, show a questionnaire for CSAM users, and analyse 11,470 responses.
65.3\% of CSAM users first saw the material when they were children themselves,
and half of the respondents first saw the material accidentally,
demonstrating the availability of CSAM\@.
48.1\% want to stop using CSAM\@.
Some seek help through Tor, and self-help websites are popular.
Our survey finds commonalities between CSAM use and addiction.
Help-seeking correlates with increasing viewing duration and frequency, depression, anxiety, self-harming thoughts, guilt, and shame.
Yet, 73.9\% of help seekers have not been able to receive it.
 \end{abstract}

\begin{document}

\flushbottom

\maketitle

\thispagestyle{plain}
\pagestyle{plain}
\pagenumbering{arabic}
\setcounter{page}{1}

\section*{Introduction}\label{sec:intro}

Society benefits from the responsible use of anonymity;
for instance, newspapers and police departments use anonymous tips as a source of information,
people in countries with strict political systems hide their identities to avoid persecution for their political views,
and individuals are permitted to speak freely about personal matters, such as religious issues.

The Onion Router (Tor) provides online anonymity for millions of internet users every day,
and it is often portrayed favourably as a method to avoid surveillance
by concealing the origin of communications,
resisting web browser fingerprinting,
circumventing traffic for unrestricted internet access without censorship,
and providing anonymous online hosting using onion domains.

On the other hand, online anonymity serves as a catalyst for the dark side of human behaviour:
it is one of the principal causes of the online disinhibition effect,
characterised by lowered psychological restraints resulting in intensified aggressive, illicit, or unethical behaviour \citep{DBLP:journals/chb/Guitton13a}
-- including higher levels of harassment, threats, racial agitation, and sexism \citep{DBLP:journals/chb/FoxCL15}.
Tor users hosting anonymous onion websites behave accordingly:
the websites predominantly host unethical or illicit content, including illegal drug trade, fraud, computer crime,
and the distribution of child sexual abuse material
(CSAM) \citep{DBLP:journals/chb/Guitton13a}.

We use the term `CSAM' instead of `child pornography' to emphasise the distinction.
The term `pornography' implies consent,
and whilst it is contested to what extent consent is present in the production and dissemination of adult pornography,
in the case of CSAM it is not possible for any child to consent in the first place.
CSAM means media, including images, videos, and live streaming, that depict sexual violence against a child.

It is common for those who search for and view CSAM to engage in other related compulsive behaviours,
such as collecting and organising CSAM by age, gender, sex act, and fantasy \citep{quayle2004child}.
In order to encompass all activities, we refer to these individuals as `CSAM users' in this article.
This group includes individuals who search for, view, disseminate, and/or trade CSAM\@.
A large portion of this group is likely to have sexual interest in children (i.e., paedophilia or hebephilia) \citep{babchishin2015online}.
These sexual preferences are classified as mental health disorders because they result in self-harm and harm to others,
and therapy can improve well-being and prevent harm to
children \citep{dsm5}.

Many users are not just passive observers of CSAM\@; rather, they are sexually motivated \citep{knack2020motivational}.
Previous research has suggested that problematic use of legal pornography can escalate to violent sexual behaviour and the use of CSAM\@.
Consumers who view legal pornography and engage in masturbation fuel this process of escalation
by providing themselves with a `powerful neurochemical reward' through orgasm \citep{sharpe2021problematic}.
This process, along with repeated exposure, may condition users into continuing to use the material despite possibly wanting to stop \citep{knack2020motivational}.

The fact that CSAM is easy to access through the Tor network -- and other anonymous networks -- increases the likelihood that more children will be sexually abused:
one study found that \rounding{41.8499353169}\% (N = 647 of 1,546) of anonymous people who answered a survey after searching for CSAM on Tor search engines
said they had tried to seek direct contact with children online after viewing CSAM,
and
\rounding{57.8913324709}\% (N = 895 of 1,546)
said they were afraid that viewing CSAM could lead to sexual acts with a child \citep{insoll2022}.
This suggests that roughly half of CSAM users do not expect to become offline offenders,
which is relevant for subsequent public health interventions,
as this may indicate a separation between the populations of online-only offenders and online and in-person offenders.

Despite abundant evidence of the growing prevalence and severe consequences of CSAM accessible through the Tor network, computer science research on CSAM remains limited.
A report to the US Congress in 2022 \citep{levine2022increasing} addresses the lack of research regarding CSAM accessible through the Tor network,
as well as:
`The ethical failure of computer science researchers with respect to acknowledging the harms against children carried out via Tor and Freenet is vast.
Dozens of papers on Tor, Freenet, and I2P have been written in the past decade and published in the flagship security
and privacy conferences of the computer science research community:
USENIX Security, ACM Computer and Communications Security, ISOC Network and Distributed Systems Security,
and the Proceedings of Privacy Enhancing Technology.
Virtually none have mentioned the harms of these anonymous services.'

All of these venues have rules for stating harms and disclosing and discussing ethical issues,
but the sponsoring organisations, chairs, and reviewers do not strictly enforce these rules \citep{levine2022increasing}.

Articles in top computer science and security venues even conduct research on Tor usage;
one even poses the research question, `Why do
people use Tor?',
and despite the fact that one of their interview responses raises the issue,
the authors make no mention of child abuse \citep{DBLP:conf/soups/GallagherPM17}.
Similarly, even when the subject of one article is sexual abuse and it references Tor Browser usage,
child abuse is not mentioned \citep{DBLP:conf/soups/Obada-ObiehSB20}.

In 2018, the USENIX Security Symposium article
`How Do Tor Users Interact With Onion Services?'
presents an online survey of 517 Tor users where several users are concerned about CSAM\@;
however, this does not lead to any further analysis in the paper, and it avoids mentioning child abuse \citep{DBLP:conf/uss/WinterERDCF18}.

Similarly, a 2019 article in the Proceedings of the Web Science Conference (WebSci) titled
`A Broad Evaluation of the Tor English Content Ecosystem'
omits any mentions of CSAM,
despite the authors' claims to have performed an exhaustive evaluation of the content and use of Tor \citep{DBLP:conf/websci/ZabihimayvanSDA19}.

There are surely articles covering the harms and studying CSAM accessible through the Tor network, but mainly outside of the top computer science venues.
In 2020, an article published in The Proceedings of the National Academy of Sciences (PNAS) acknowledged
that `The Tor anonymity network allows users to protect their privacy and circumvent censorship restrictions
but also shields those distributing child abuse content' \citep{doi:10.1073/pnas.2011893117}.

As early as 2011, research revealed the widespread distribution of CSAM in peer-to-peer networks \citep{aked2011investigation}.
This issue affects not only Tor but also the I2P and Freenet anonymity networks;
in a 2022 investigation, a Freenet content analysis revealed that \rounding{11.95}\% of the 7,161 analysed freesites contained CSAM \citep{DBLP:journals/istr/Figueras-Martin22}.
The first systematic analysis in 2013 indicated that \rounding{17.5918018787}\% (206 of 1,171) of the onion services surveyed shared CSAM
and concluded `The support for the further development of Tor hidden services should hence stop' \citep{DBLP:journals/chb/Guitton13a}.
Measurements on onion service visits find CSAM sites to be the most popular ones \citep{DBLP:journals/iet-ifs/OwenS16}.
In 2014, an estimated 17\%
of the onion websites provided sexual material, of which about half was CSAM \citep{DBLP:conf/isi/SpittersVS14}.
In 2018, additional research pointed in the same direction \citep{DBLP:journals/di/DalinsWC18}.

We collect and analyse web content accessible through onion websites, study user searches on the Tor search engine, \AHMIA{},
and demonstrate that CSAM is widely available and Tor users actively seek this material.
We show a questionnaire for those who search for CSAM and analyse 11,470 responses.

Our research questions (RQs) are:
\begin{description}
    \item[\textbf{\autoref{rq:stats}}.]\refstepcounter{rqcounter}\label{rq:stats}
    What is the distribution volume of CSAM hosted through the Tor network?
    \item[\textbf{\autoref{rq:searches}}.]\refstepcounter{rqcounter}\label{rq:searches}
    What is the CSAM search volume, and what exactly are users seeking?
    \item[\textbf{\autoref{rq:questionnaire}}.]\refstepcounter{rqcounter}\label{rq:questionnaire}
    What does the survey reveal about CSAM users?
    \item[\textbf{\autoref{rq:demo}}.]\refstepcounter{rqcounter}\label{rq:demo}
    How can search engine-based intervention reduce child abuse?
\end{description}

\begin{figure*}[ht!]
\centering
\ifLINUXBUILD%
\includegraphics[width=\textwidth]{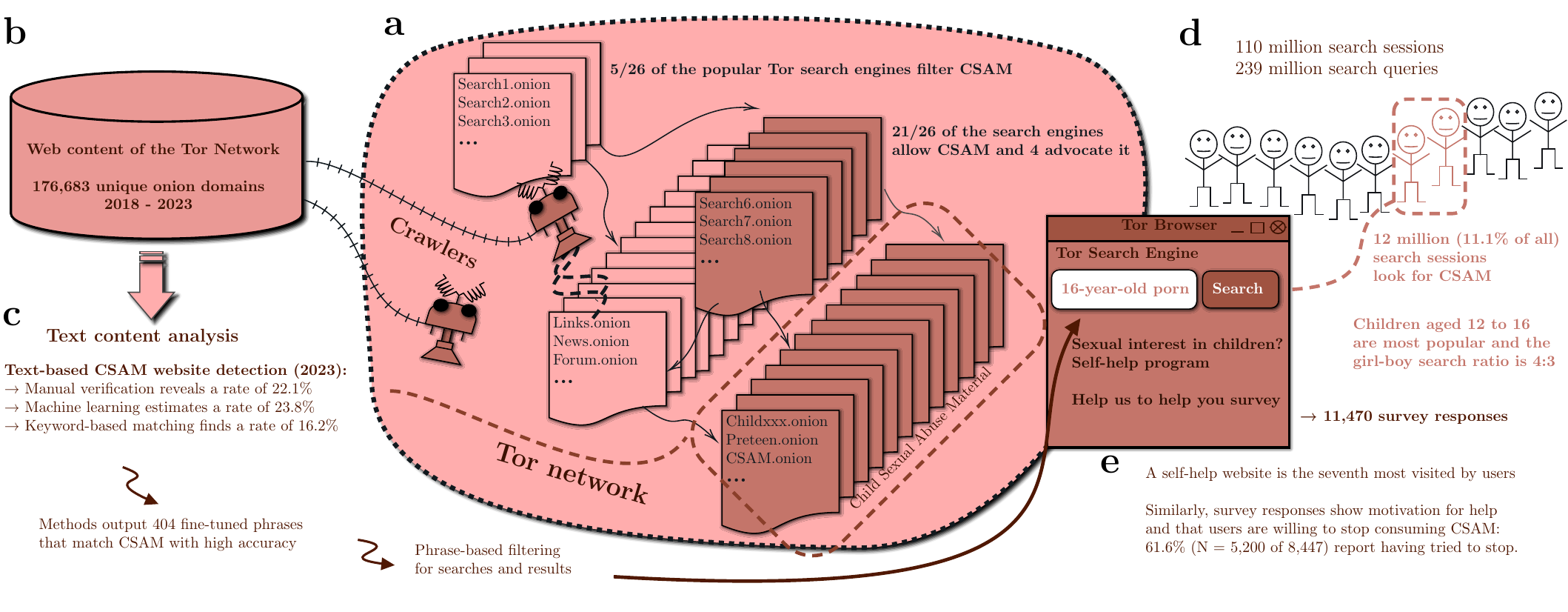}
\else
\Huge\textbf{PLACE\\HOLDER}
\fi
\caption{
We investigate the availability of CSAM hosted through the Tor network and its users.
\underline{(a)} Tor enables anonymous web publishing through onion domains.
These websites host a variety of content, and there are Tor-specific search engines for searching.
21/26 of the popular Tor search engines allow CSAM websites.
\underline{(b)} Our web crawlers collected online content from 176,683 different onion domains from 2018 to 2023.
\underline{(c)} Using text-based CSAM detection methods, we investigate the number of websites sharing CSAM\@.
The identification methods provide us with 404 phrases that accurately identify CSAM content.
\underline{(d)} This enables text-based detection and filtering for Tor search engines.
CSAM-related searches are among the most popular of the 239 million total queries.
Out of 110 million search sessions, 11.1\% are seeking CSAM\@.
\underline{(e)} The search engine directs CSAM-seekers to self-help websites and asks them to complete our survey.
The results indicate that they want assistance and are motivated to stop using CSAM\@.
}\label{fig:results}
\end{figure*}

\autoref{fig:results} shows our approaches for analysing CSAM availability and usage on Tor.
The Methods section presents our research methodology in detail.
Our contributions are:
(1) We measure the CSAM distribution hosted through the Tor network over a five-year period using onion website crawling,
which indicates that through the years 2018--2023 about one-fifth of the websites share CSAM\@.
(2) We show that these CSAM websites can be reached directly from the majority (21/26) of the top Tor search engines used today,
and four Tor search engines even advertise CSAM\@.
(3) We find that \rounding{11.1410}\% (N = 12,270,042 of 110,133,715) of the search sessions are explicitly searching for CSAM\@;
the single phrase `child porn' is one of the top queries.
(4) When we prompt those who search for CSAM with a survey,
we find that \rounding{61.560317}\% (N = 5,200 of 8,447 who replied to the question) of CSAM users have tried to stop watching CSAM,
\rounding{48.097128}\% (N = 4,120 of 8,566 who replied to the question) want to stop using CSAM,
and there is an unmet demand for help resources.
(5) Search engines are a key part of the solution;
hence, we demonstrate the effectiveness of CSAM detection and the ability of search engines to intervene and steer CSAM users towards help.

\section*{Results}\label{sec:results}

\subsection*{Surge of CSAM hosted through the Tor network}\label{sec:stats}

\textit{\autoref*{rq:stats}: What is the distribution volume of CSAM hosted through the Tor network?}

\begin{figure*}[ht!]
\centering
\ifLINUXBUILD%
\includegraphics[width=1.0\textwidth]{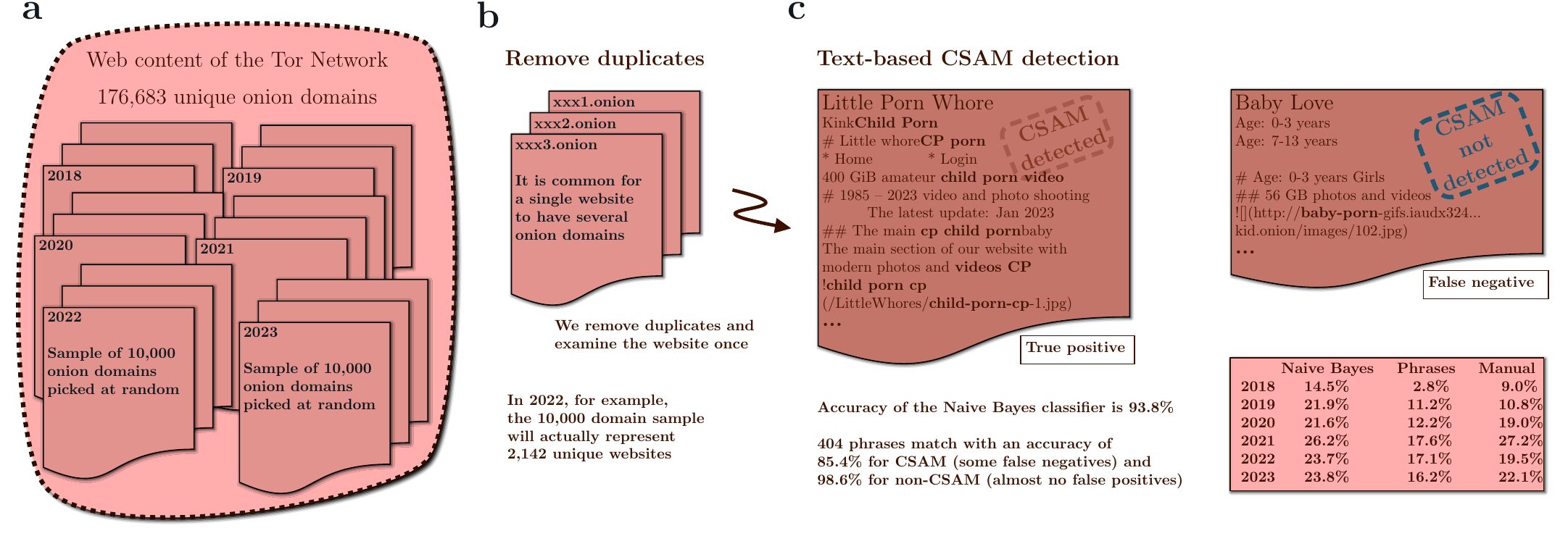}
\else
\Huge\textbf{PLACE\\HOLDER}
\fi
\caption{
We measure the proportion of CSAM onion websites inside the Tor network in 2018--2023.
\underline{(a)} We use a sample of 10,000 unique onion domains for each year.
\underline{(b)} Many websites have several onion domains.
We compare the title and sentences of the pages to detect duplicates, and restrict to a single domain if multiple domains share identical content.
\underline{(c)} We execute text-based CSAM detections against the content of these distinct domains.
Some CSAM websites do not use explicit sexual language, and text-based detection fails.
Using three separate methods -- manual validation, phrase matching, and the naive Bayes classifier -- we
discover that the detected percentage of websites sharing CSAM is 16.2--23.8\% in 2023.
Comparing automated methods to human validation by hand yields consistent results (22.1\% in 2023 and 19.5\% in 2022).
We randomly select the plain text representations of 1,000 onion websites for each year, 2018--2023,
and read the text content of these websites to determine whether they share CSAM and what the English vocabulary
is for this type of page.
}\label{fig:detection}
\end{figure*}

We investigate the years 2018--2023 and use a sample of 10,000 unique onion domains for each year.
We then subject the text content to duplicate content filtering, fine-tuned phrase search, and supervised learning classifiers.
This returns the detected CSAM percentage for each year, as shown in \autoref{fig:detection}.

The phrase matching fails to detect anything that does not describe CSAM with the obvious phrases.
Furthermore, there are real CSAM websites that do not use explicit sexual language and instead refer to content such as `baby love videos'.
Our text-based detection does not match these websites.
In addition, there are link directory websites that provide descriptions of CSAM website links.
Our text-based detection matches the description phrases despite the fact that this type of website does not share CSAM, rather merely links to websites that do.

We achieve \rounding{93.8}\% accuracy with a basic naive Bayes classifier (see Supplementary Methods A.2).
Some legal adult pornography websites, like \emph{PornHub}, provide an alternative onion domain accessible via Tor.
We manually review a sample set of \emph{PornHub} pages, and there is no indication of anything other than adult material;
therefore, we include these in our training data to teach the classifier to differentiate between legal and illegal content.
The classifier performs well and can distinguish between legitimate pornography websites and unlawful CSAM websites.
The accuracy is as expected and actually quite consistent with previous research (\rounding{93.5}\%) for Tor content classification \citep{he2019classification}.

We anticipate that the phrase matching method will generate few false positives due to the explicit nature of the matching phrases.
It is rare for a website to contain these phrases unless it also contains CSAM, and even those exceptions are describing linked CSAM websites.
We anticipate -- for the same reason --
that this method will generate false negatives, as it requires exact CSAM-describing language.
Indeed, the matching works accordingly:
its accuracy is \rounding{85.35}\% with CSAM websites (some false negatives) and \rounding{98.60}\% with non-CSAM websites (almost no false positives).

When law enforcement has seized control of CSAM servers operating through the Tor network,
they have documented terabytes of content with hundreds of thousands of users (see Supplementary Discussion C.2).
In a comparable manner, we find indicators that suggest the distribution of extensive CSAM collections.
While we read texts from the websites,
we see numerous CSAM websites that claim to share gigabytes of media and thousands of videos and images.

Using three separate methods -- manual validation, phrase matching, and the naive Bayes classifier -- we
conclude that around one-fifth of the unique websites hosted through the Tor network share CSAM\@.
Previous research, in 2013, found that \rounding{17.5918018787}\% of onion services shared CSAM, which corresponds with our findings \citep{DBLP:journals/chb/Guitton13a}.

\subsection*{Examining CSAM user behaviour}\label{sec:searches}

\textit{\autoref*{rq:searches}: What is the CSAM search volume, and what exactly are users seeking?}

\subsubsection*{11.1\% of the search sessions seek CSAM}

We examine search chains generated by users who enter consecutive queries.
We follow the searches entered by users, track queries per user, and study 110,133,715 total search sessions,
and discover that \rounding{32.46201}\% (N = 35,751,619) include sexual phrases.
Finally, we find that \rounding{11.1410}\% (N = 12,270,042) of the search sessions reveal
that the user is explicitly searching for content related to the sexual abuse of children.
Some of these CSAM search sessions include either `girl\@(s)' (393,261) or `boy\@(s)' (289,407);
searching for girls is more prevalent, with a ratio of 4:3.
Seeking torture material is not typical:
\rounding{0.4680}\% of CSAM search sessions (57,429) contain the terms `pain', `hurt', `torture', `violence', `violent', `destruction', or `destroy'.

During the COVID-19 pandemic and the first months of lockdowns, there was a significant surge in the user base of legal pornography websites across nations \citep{harris2021child}.
Surprisingly, we find that before and after COVID-19 pandemic measures (lockdowns, individuals spending more time at home),
there were no significant changes in the behaviour of CSAM users (see more in Supplementary Methods A.8).

\subsubsection*{54.5\% are searching for 12- to 16-year-olds}

\begin{figure}[ht!]
  \begin{minipage}[b]{0.5\linewidth}
    \centering
    \ifLINUXBUILD%
    \includegraphics[width=1.0\linewidth]{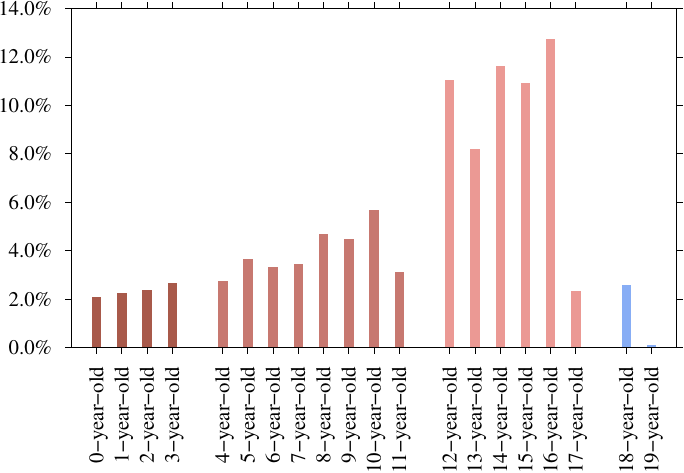}
    \else
    \Huge\textbf{PLACE\\HOLDER}
    \fi
    \caption{Ages between zero and 17 included in the CSAM search sessions and search sessions seeking 18-year-old and 19-year-old adults as a comparison (N = 479,555).
    16-year-old (N = 61,083 of 479,555 -- 12.7\%) is the top-mentioned age.
    54.5\% of age-revealing searches (N = 261,162 of 479,555) target those aged 12--16 years old.
    Outside of this age range, the interest declines.
    }\label{fig:ages}
  \end{minipage}
  \hspace{0.2cm} %
  \begin{minipage}[b]{0.5\linewidth}
    \centering
    \ifLINUXBUILD%
    \includegraphics[width=1.0\linewidth]{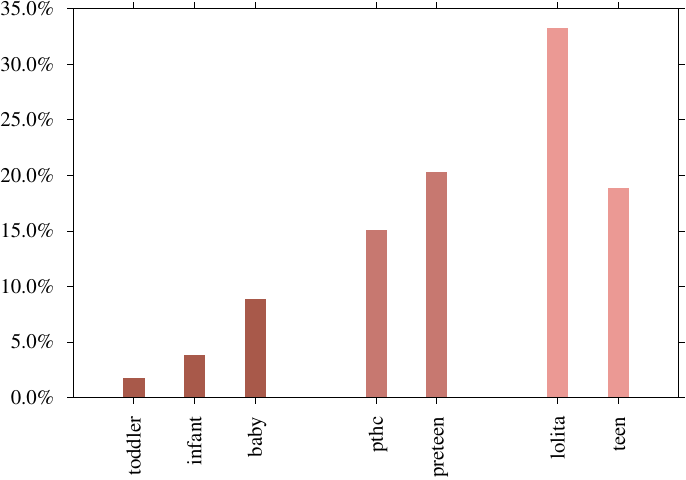}
    \else
    \Huge\textbf{PLACE\\HOLDER}
    \fi
    \caption{
    In the context of explicit CSAM search sessions, there are a total of 2,287,057 broad age-indicating searches with terms
    `toddler', `infant', `baby', `pthc' (preteen hardcore), `preteen' (preadolescence, ages between nine and 12),
    `lolita' (refers to a girl around 12--14 years old),
    and `teen' (when included with CSAM terms).}\label{fig:agewords}
  \end{minipage}
\end{figure}

We determine if the search session reveals the exact age in which the CSAM user is interested.
For example, for a 13-year-old, we count search sessions that include 13y\@(*), 13+y\@(*), 13teen, thirteen+year, 13boy\@(s) or 13girl\@(s).
We use the same logic for other ages.
In addition, to compare these searches for CSAM to legal adult sexual content searches,
we include search sessions seeking 18-year-old (N = 12,347 from 110,133,715) and 19-year-old adults (N = 458 from 110,133,715).
We find the age information in total for 479,555 search sessions.
\autoref{fig:ages} illustrates the age distribution of CSAM queries.
This age distribution aligns with findings from previous studies \citep{steel2021collecting}.
An article titled `Pedophilia, Hebephilia, and the DSM-V' finds qualitative differences between offenders
who preferred pubertal and those with a prepubertal preference (a clinical trial of 881 men with problematic sexual behavior) \citep{blanchard2009pedophilia}.
The authors also note that the majority of child abuse victims are 14 years old.
They concluded that the psychiatric diagnoses should be improved to include the following:
sexually attracted to children younger than 11 (paedophilic type),
sexually attracted to children aged 11--14 (hebephilic type),
or sexually attracted to both (pedohebephilic type).
Our data suggests that it may be more appropriate to observe the high percentage of individuals who have a sexual interest in 12-year-olds but not 11-year-olds.
This finding is consistent with the national average age at menarche in the United States, which is 12.54 years \citep{anderson2003relative}.
Additionally, observe the decline in sexual interest that occurs across the ages of 17, 18, and 19, which indicates a distinct sexual interest in those aged 12 to 16 years old.

Moreover, in \autoref{fig:agewords}, we investigate CSAM search sessions containing age-indicating search terms
and find that users are predominantly interested in 12- to 14-year-old teen content; for example,
`lolita'
is the most popular term when
compared to other age-related terms,
with 33.2\% (N = 746,786 of 2,287,057).
In Vladimir Nabokov's 1955 novel `Lolita',
a middle-aged male is sexually attracted to a 12-year-old girl and sexually abuses her. %
In the 1962 Stanley Kubrick film adaptation of the novel, `Lolita' is 14 years old.

\subsubsection*{How accessible is CSAM using Tor?}

Based on result click statistics, we rank the top 26 most visited search engines online on 17 March 2023.
To determine whether CSAM is permitted in search results,
we test the searches `child', `sex', `videos', `love', and `cute', then study the search results.
21 out of 26 search engines provide CSAM results.
Four search engines even promote and advocate CSAM\@.
One of them even states that `child porn' is the number one search phrase.
It is positive that five Tor search engines attempt to block CSAM\@.
Yet, a user can utilise these search engines to locate other search engines and ultimately locate CSAM through the latter.
Even if search engines block sites that directly share CSAM, it is still possible to find other entry points for onion sites that provide links to CSAM websites.
With any major Tor website entrypoint, search engine, or link directory, a Tor user is only a few clicks away from CSAM content.

\subsubsection*{Self-help services are reaching users}

When we study the searches, we discover that there are a few hundred queries
from people who want to cease viewing CSAM and are concerned about their sexual interest in children,
including queries
`overcome child porn addiction' and
`how to stop watching child porn' (see more in Supplementary Methods A.6).

When a person searches for CSAM, three prominent Tor search engines
provide only links to self-help programmes for those who are
concerned about their thoughts, feelings, or behaviours.
The intervention of CSAM searches directs individuals away from CSAM and towards help.
Data from one of the self-help websites indicates that CSAM users actively visit the website,
and those who start the self-help programme are very likely to continue following the programme (see more in Supplementary Methods A.6).
In the next section, we show that when we present a survey for those who search for CSAM,
many reply with motivation to stop using CSAM\@.
\subsection*{Intervention for CSAM users}\label{sec:intervention}

\textit{\autoref*{rq:questionnaire}: What does the survey reveal about CSAM users?}

\begin{figure*}[ht!]
\centering
\ifLINUXBUILD%
\includegraphics[width=1.0\textwidth]{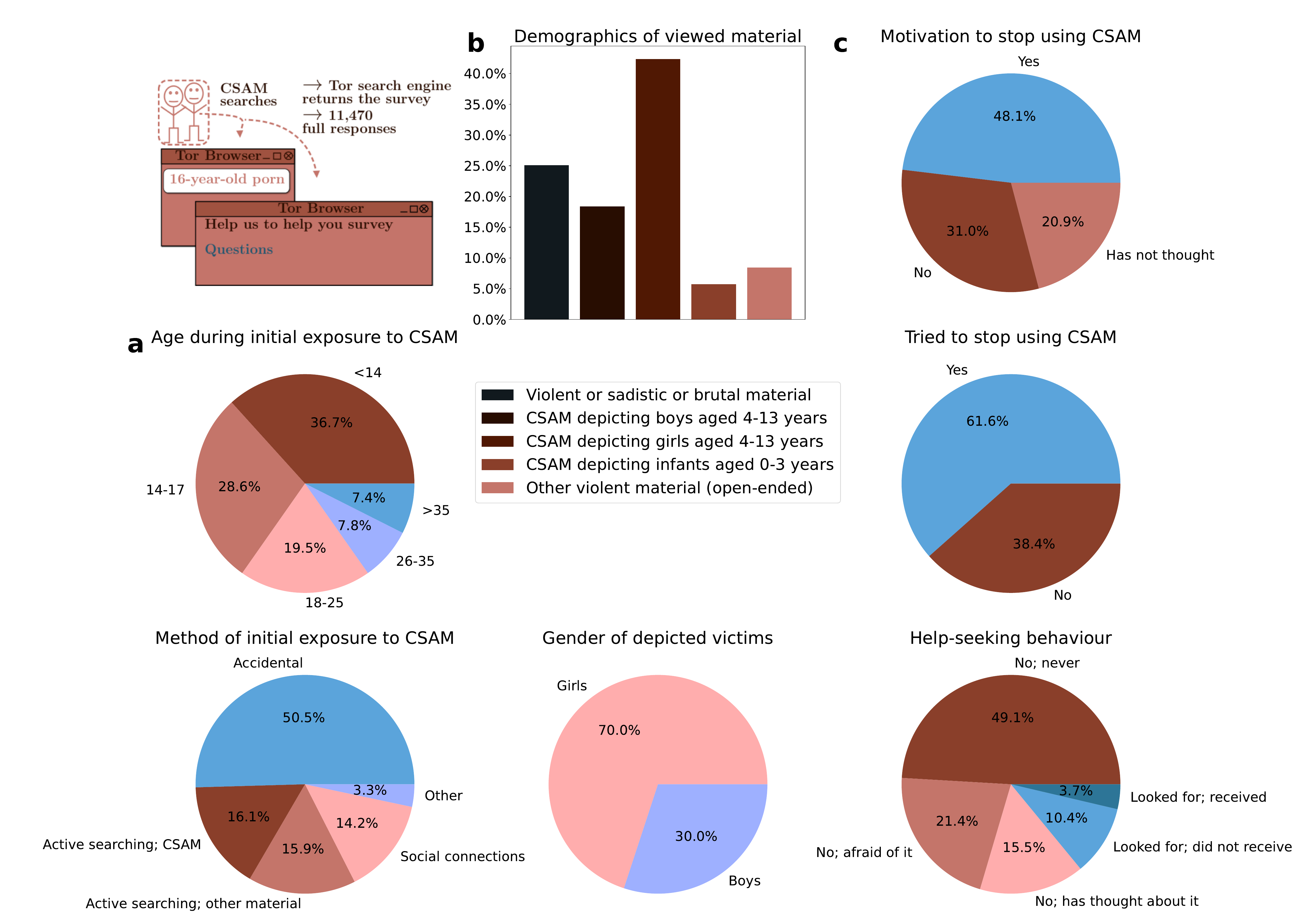}
\else
\Huge\textbf{PLACE\\HOLDER}
\fi
\caption{
Our anonymous survey received responses from 11,470 individuals who sought CSAM on three popular Tor search engines.
\underline{(a)} The survey results indicate that 65.3\% (N = 7,199 of 11,030 who replied to the question) of CSAM users first saw the material when they were under 18 years old.
36.7\% (N = 4,048 of 11,030) first saw CSAM when they were 13 years old or younger.
50.5\% (N = 4,843 of 9,599 who replied to the question) report that they first saw CSAM accidentally.
\underline{(b)} We asked the respondents what types of images and videos they view.
Viewing CSAM depicting girls is more prevalent, with a ratio of 7:3.
\underline{(c)} The survey results indicate that 48.1\% (N = 4,120 of 8,566 who replied to the question) of CSAM users are willing to change their behaviour to stop using CSAM,
and 61.6\% (5,200 of 8,447 who replied to the question) have tried to stop using CSAM\@.
However, only 14.0\% (N = 985 of 7,013 who replied to the question) of CSAM users have sought help to stop using CSAM,
and an even smaller portion of 3.7\% (N = 257) have actually received help.
Notably, 21.4\% (N = 1,498) are afraid to seek help.
}\label{fig:survey}
\end{figure*}

Individuals who seek CSAM on Tor search engines answered our survey
that aims at developing a self-help programme for them.
\autoref{fig:survey} aggregates statistics from the responses.

Most CSAM users were first exposed to CSAM while they were children themselves,
and half of the respondents (N = 4,843 of 9,599 who replied to the question) first saw the material accidentally,
demonstrating the accessibility and availability of CSAM online.
Exposure to sexually explicit material in childhood is associated,
inter alia, with risky sexual behaviour in adulthood \citep{lin2020exposure},
sexual harassment perpetration \citep{brown2009x},
and the normalisation of violent sexual behaviour \citep{hunter2010developmental},
and has been defined as an adverse childhood experience and a form of noncontact sexual abuse \citep{asmussen2020adverse}.

We ask for information regarding two age ranges in the survey: zero to three years and four to 13 years.
We structured the question with the intention of focusing on pre-pubescent children, aged 0--13.
Respondents were able to specify the age in the option `Other violent material, what?'
The majority (\rounding{60.73215}\%, N = 5,342 of 8,796 who replied to the question) of respondents say they view CSAM depicting girls or boys aged between four and 13 years,
indicating a preference for images and videos depicting prepubescent and pubescent children.
Of the respondents, \rounding{69.730438}\% (N = 3,725 of 5,342) say they view girls,
compared to \rounding{30.26956}\% (N = 1,617 of 5,342) who view boys.
There is also a small group \rounding{5.75261}\% (N = 506) of CSAM users with a preference for CSAM depicting infants and toddlers aged between zero and three years old.
Additionally, \rounding{25.06821}\% (N = 2,205) reported viewing images and videos related to violent or sadistic and brutal material.

\vspace{5pt}
\vspace{5pt}

\noindent %
\begin{minipage}[c]{0.45\textwidth}
  \rounding{8.44702}\% (N = 743) of respondents say they view `other violent material',
  and 458 provide explanatory open-ended responses.
  \rounding{61.572052}\% (N = 282) of these responses explicitly mention the age of children depicted in the CSAM viewed,
  providing N = 1,637 mentions of age, as \autoref{fig:survey_ages} illustrates.
  Most responses refer to age brackets, for example `over 12 years old', in which we define this to mean 12--17.
  The most common age is 15-year-old (N = 234), followed by 16-year-old (N = 221), and 14-year-old (N = 209).

  The survey data provides an age distribution that appears similar but is distinct from the search data (see statistical tests in Supplementary Methods A.5).
  The survey responses give the age ranges that respondents say they are interested in
  -- whereas the search sessions reveal the precise age that people are most interested in.
  \rounding{54.4592382521}\% of age-revealing CSAM search sessions target sexual content aimed at 12- to 16-year-olds.
  This survey's age distribution yields an almost identical percentage:
  the range between 12 and 16 years old accounts for
  \rounding{56.7501527}\% (N = 929 mention age in this range from all 1,637 mentioned ages).
\end{minipage}%
\hfill %
\begin{minipage}[c]{0.5\textwidth}
  \includegraphics[width=\linewidth]{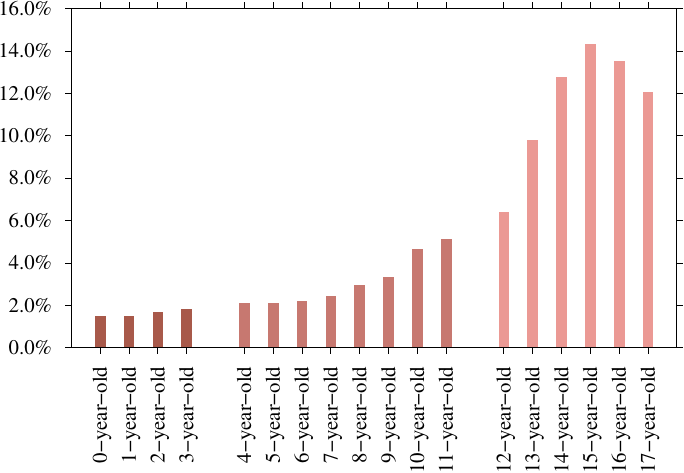}
  \captionof{figure}{282 of the open-ended responses explicitly mention the age of children depicted in the CSAM viewed, providing 1,637 mentions of age.}\label{fig:survey_ages}
\end{minipage}

\vspace{5pt}

Our analysis of 458 open-ended responses for `Other violent material, what?' supports the prevalence of viewing material depicting girls.
\rounding{33.84279}\% (N = 155) of the 458 open-ended responses explicitly mention the gender of children.
\rounding{91.612903}\% (N = 142) of the responses that mention gender refer to girls,
and \rounding{30.32258}\% (N = 47) refer to boys.
\rounding{21.93548}\% (N = 34) of the responses mention both girls and boys.
Considering the quantitative and qualitative data together (N = 5,488),
viewing CSAM depicting girls (N = 3,839, \rounding{69.95262}\%) is more prevalent than viewing CSAM depicting boys (N = 1,649, \rounding{30.047376}\%),
with a ratio of 7:3.

Overall, these findings are consistent with the latest Internet Watch Foundation's Annual Report 2022 (see Supplementary Methods A.9). %

\subsubsection*{CSAM users do want assistance}

The Prevention Project Dunkelfeld offers clinical and support services to men who experience
sexual attraction towards children and reaches these individuals with media campaigns \citep{beier2009can}.
Similarly, in collaboration with some legal adult pornography websites, including \emph{Pornhub},
the \emph{Stop It Now!} organisation alerts users who conduct CSAM searches of the illegality of their actions and directs them to help resources \citep{harris2021child}.

Our results verify the feasibility of this type of intervention: a large proportion of CSAM users report that they want and have tried to change their behaviour to stop using CSAM\@.
Almost half of the respondents report wanting to stop viewing CSAM monthly, weekly, or nearly every time (\rounding{48.097128}\%,
N = 4,120 of 8,566 who replied to the question), and
the majority of respondents report having tried to stop
(\rounding{61.560317}\%, N = 5,200 of 8,447 who replied to the question).
\rounding{31.00630}\% (N = 2,656) say that they do not want to stop,
and \rounding{20.8965678}\% (N = 1,790) say they have not thought about it.

Despite self-reported willingness and attempts to change behaviour,
the fact that they are responding to the survey is evidence of their continued search for CSAM
-- albeit temporarily stepping away from CSAM to contemplate the concerns posed in the survey and provide a response.
This raises the question of the commonalities between CSAM use and addiction.
While addiction to the internet is not listed as a diagnostic disorder in the
\emph{Diagnostic And Statistical Manual Of Mental Disorders, Fifth Edition, Text Revision (DSM-5-TR)} \citep{dsm5},
there is extensive debate over whether problematic use of the internet
-- in particular in the context of legal pornography and CSAM use --
can be considered an addiction.
\emph{The International Classification of Diseases 11th Revision (ICD-11)}
includes compulsive sexual behaviour disorder,
characterised by `persistent pattern of failure to control intense, repetitive sexual impulses or urges resulting in repetitive sexual behaviour'
-- including repetitive use of legal pornography \citep{icd11}.

Repetitive pornography use has similar effects to substance addiction \citep{laier2013cybersex}.
Many people continually use CSAM and display addictive behaviours \citep{quayle2003model},
and the intensity of CSAM use often has properties that users call addictive \citep{quayle2006sex}.
In the search engine data, we notice that users who seek help often refer to their condition as `child porn addiction'.
Understanding the commonalities between CSAM use and addiction is beneficial to prevention and treatment.

\subsubsection*{Help-seeking behaviour among CSAM users}

Despite many respondents reporting that they would like to stop using CSAM,
help-seeking behaviour among CSAM users remains low.
Only \rounding{14.0453443605}\% (N = 985 of 7,013 who responded to the question) of CSAM users have sought help.
Many report that they feel afraid to seek help (\rounding{21.4}\%, N = 1,498 of 7,013),
and the majority \rounding{64.6}\% (N = 4,530 of 7,013) report that they have not sought help.

Of those who have actively sought help to change their illicit behaviour,
\rounding{73.908629}\% (N = 728 of 985) have not been able to get help.
This population has an unmet demand for effective intervention resources \citep{levenson2017obstacles}.
This may be due to a lack of awareness of the resources available \citep{schmidt2022outpatient}
or because the available resources are not desirable.
Recent studies \citep{levenson2017obstacles,schmidt2022outpatient,grady2019can}
found the following barriers to seeking and receiving psychological services for child sexual offenders and people concerned about their sexual interest in children:
fear of legal consequences;
fear of stigmatisation;
shame;
affordability;
and a perceived or actual lack of understanding by professionals.

The unmet demand for help resources demonstrates the urgent need for investment
and further implementation of perpetration prevention programmes in order to effectively reach those who require intervention \citep{letourneau2014need}.

Through bivariate analysis,
we examine the associations between a number of covariates and the outcome of help-seeking in the survey data.
We find the following to be determinants of help-seeking:
duration and frequency of CSAM use;
depression, anxiety, self-harming thoughts, guilt, and shame.
We have only included the respondents with non-missing answers;
for all results on determinants of help-seeking, see Supplementary Methods A.4 and Supplementary Tables.

Individuals who have used CSAM for a longer duration and those who use it more frequently are more likely to actively seek assistance to change their behaviour.
There is an opportunity to intervene with those who have been viewing CSAM for a shorter duration in order to increase help-seeking at an earlier stage.
Those who use CSAM more frequently may be more likely to seek help due to the detrimental impact that frequent use of CSAM may have on an individual's daily life,
including impaired social and occupational functioning and deep distress,
which may be a strong motivator to seek help to reduce use of CSAM in order to improve life situations.
Common reasons for seeking help for substance abuse include habitual use,
taking a substance for a long time, and a need to take it daily \citep{nebhinani2012reasons}.
Such driving factors for help-seeking appear to be similar in this sample of CSAM users.
Respondents who face more difficulties in carrying out ordinary daily routines and activities are more likely to have sought help to stop using CSAM\@.
Those who experience such difficulties daily have one of the highest rates of help-seeking.

\textit{\autoref*{rq:demo}: How can search engine-based intervention reduce child abuse?}

We demonstrate that not only are CSAM websites widely hosted through the Tor network, but that they are also actively sought.
However, in contrast, instead of watching CSAM, individuals voluntarily participate in the search engine-prompted survey. Consequently, even this intervention reduces CSAM usage.

We propose an intervention strategy based on our observation that some CSAM users do indeed recognise their problem.
Even when CSAM users are seeking CSAM, they are willingly visiting self-help pages and continuing to study cognitive behavioural therapy information.
Search engines, which are the main way people find onion sites, should start filtering CSAM and diverting people towards help to stop seeking CSAM\@.
This is technically possible because of the accurate, text-based detection of CSAM that we demonstrate, and furthermore, the CSAM phrase detection list can be shared between search engines.

\section*{Discussion}\label{sec:discussion}

\subsection*{Lack of interdisciplinary research}
Technical and non-technical scientists work in separate silos and publish in separate venues,
and these venues -- including their peer-review processes -- promote these silos by focusing on technical or non-technical research.
In the present era of 2024
-- when the online environment is common and data facilitates innovative research --
psychology journals display a hesitancy to publish articles that employ technical methodologies (such as a Naive Bayes classifier).
This leads to a lack of an overall methodology to seek solutions to reduce child abuse and CSAM\@.
Interdisciplinary research provides key insights to our work:
we combine survey methods and social scientists with computer scientists to produce holistic research instead of fragmented views.

\subsection*{Unwillingness to acknowledge CSAM in the top-ranked computer science venues}
How is it possible that there are so many studies classifying Tor websites and usage without addressing CSAM\@?
A plausible explanation is that the researchers omitted the CSAM findings
from their data without providing an explanation in the articles (i.e., `A Broad Evaluation of the Tor English Content Ecosystem' \citep{DBLP:conf/websci/ZabihimayvanSDA19}).

We find a big gap in relevant research in the top-ranked venues in the field of computer science.
Prior investigations encompassed a limited number of scholarly articles of comparable calibre, and these yielded consistent results with our own:
in 2013, a systematic analysis determined that 17.6\% of onion websites distribute CSAM \citep{DBLP:journals/chb/Guitton13a}, and in 2016,
research revealed that CSAM websites are the most popular among Tor users \citep{DBLP:journals/iet-ifs/OwenS16}.
However, despite these findings, computer scientists have continued to neglect CSAM distribution through the Tor network.
Could this be due to the contentious nature of CSAM\@?

Computer scientists should evolve Tor and other anonymity networks
so that the privacy goals are consistent with ethical and legal concerns \citep{collier22,sarda2023onion}.
Current peer-to-peer networks have essentially no remedies for widespread abuse;
this is a problem that we intend to investigate in our future research.

\subsection*{Policy to combat CSAM and implement public health programmes for CSAM users}
CSAM provides a paramount example of how technology can be used in harmful ways.
As highlighted in the ninth report about model legislation and global review by
the International Centre for Missing and Exploited Children \citep{icme2018}, a
global effort must be further conducted to harmonise the legal and regulatory
framework in the international arena.

Policies aimed at preventing the spread and use of online CSAM should be implemented.
Situational Crime Prevention (SCP) is a criminological approach that employs five strategies and 25 techniques to reduce crime opportunities \citep{clarke1995situational}.
SCP has been shown to reduce criminal behaviour \citep{HO2022102611},
and these tactics might be effective against CSAM \citep{doi:10.1080/13552600.2019.1705925}.
As an illustration, a study examining individuals who accessed a honeypot website that displayed pornography portraying adult actresses as children
found that online warning messages offer an effective and scalable tactic to reduce access to CSAM \citep{prichard2022effects}.

It is urgent to deploy public health programmes for CSAM users.
These individuals are motivated to seek help, but the help is largely currently unavailable.
There is a growing global epidemic of CSAM usage, and some describe masturbation and pornography
as coping mechanisms to alleviate economic strain, feelings of isolation, depression, and anxiety \citep{harris2021child}.
Such public health prevention programmes should be initiated, financed, evaluated, and developed not by a single industry or actor but as part of a holistic approach.
It must be the task of a broad range of actors to take responsibility for the prevention of sexual violence against children,
including but not limited to governments, the technology industry, international organisations, and civil society.

Established in Germany, the Prevention Project Dunkelfeld offers cognitive behaviour therapy to improve coping skills, stress management,
and control sexual attraction towards children \citep{beier2009can}.
The impact assessment of the `Stop It Now!' campaign demonstrates the high effectiveness of the public health approach in preventing child sexual abuse.
A series of awareness-raising films widely disseminated through media channels can successfully reach people concerned about their or others' behaviour, directing them to help services.
After establishing trust and committing to treatment, individuals who are sexually attracted to children can gain the ability to consistently regulate their impulses.
The favourable confidentiality legislation in Germany, which prohibits therapists from disclosing planned or actual child abuse offences, naturally strengthens this trust.
Project evaluation shows that post-treatment recidivism is lower among individuals who commit contact offences as opposed to child sexual abuse material users.
This demonstrates the need to develop tailored interventions based on the offender's background and behaviour.

Previous research indicates that anonymous online therapy reduces the use of CSAM\@:
the Prevent It study, a clinical trial of an online therapist-supported cognitive behavioural therapy,
indicates promising results in terms of the feasibility of dark web recruitment
and the effectiveness of anonymous online interventions \citep{latth2022effects}.
These public health programmes should offer in-person psychotherapy,
anonymous online self-help material in all languages,
anonymous online person-to-person support, and an overall drive towards treatment.

\section*{Methods}\label{sec:methods}

\subsection*{Participants}

The study and its methods are in accordance with relevant guidelines and regulations.
The Board of Suojellaan Lapsia, Protect Children ry.\ approved the experimental protocols with human participants,
in accordance with the Declaration of Helsinki Ethical Principles for Medical Research Involving Human Subjects:
(i) All participants in the \emph{Help us to help you} survey have provided informed consent.
(ii) These participants received clear information on the purposes of the study before beginning the survey.
(iii) Without compensation, they volunteered to respond to the questions.
(iv) No personal or identifiable information was collected.
(v) The survey data is stored and managed exclusively by the research team at Suojellaan Lapsia, Protect Children ry.\ without anyone else
-- not even the co-authors -- having access to the collected survey answers.

\subsection*{Intervention for CSAM users}

Previous research discovers how people find onion websites \citep{DBLP:conf/uss/WinterERDCF18}:
`The three most popular ways that almost half of our survey participants discovered onion sites by were via
(i) social networking sites such as Twitter and Reddit (48\%),
(ii) search engines such as Ahmia, (46\%) and
(iii) randomly encountering links when browsing the Web (46\%).'
Since Tor search engines serve as popular entry points to CSAM, we requested that search engines recruit Tor users who access CSAM to answer our survey.

Three prominent Tor search engines -- \AHMIA{}, OnionLand, and Onion Search Engine -- display our questionnaire to the user who searched for CSAM\@.
In this research, we analyse the responses of users who searched for CSAM on Tor web search engines using at least one of the 179 search phrases used to find CSAM\@.
The search phrases in English, Russian, Spanish, and Swedish are only used to locate CSAM, e.g., the term `childporn'.
When a user submits a query containing any of these terms on one of these three Tor search engines,
they are instead given the opportunity to voluntarily participate in the survey, which is available in 21 languages.

We may potentially be targeting a specific population due to the fact that the demographics of Tor users are probably not representative of all internet users.
Furthermore, there is a possibility that the English-speaking population is overrepresented,
as users who conduct their initial search in English may have limited vocabulary and thus be unable to identify our survey invitation in order to continue responding.

The participants in the sample are Tor users who (i) conducted a search for CSAM and (ii) opted to complete the survey; thus, they constitute a convenience sample.
Although the sample is informative, it does not generalise to all CSAM users, and there is a high probability of selection effects at play.
The absence of identifying information in the survey permits multiple responses from a single respondent.
The trend of decreasing new responses over time suggests that users who have previously encountered the survey are less likely to respond to it.

The \emph{Help us to help you} survey consists of 32 questions,
takes about 15 to 20 minutes to complete, and participants receive no compensation.
For this study, we analyse responses to 12 survey questions: 1, 2, 3, 4, 5, 7, 8, 13, 20, 22, 24, and 28 (see Supplementary Tables B).

The survey does not request any personally identifiable information from respondents --
such as age, country, or gender -- that would put privacy at risk.
Questions avoid specifics of criminal conduct (e.g., time, date, place, or victim details).
We ask CSAM users about their thoughts, feelings,
and actions related to their use of CSAM so that in the future we can build a cognitive behavioural theory-based anonymous rehabilitation programme for CSAM users.

We included the term `illegal violent material' for those respondents who do not categorise the material they view as CSAM
but indicated to us via their search terms that they are in fact searching for material depicting sexual abuse of children.

We analyse responses from all participants who answered our \emph{Help us to help you} survey from 5 May 2021 to 28 February 2023 (N = 11,470)
and compare the tendencies and habits of people who searched for CSAM (see Supplementary Methods A.4 and Supplementary Tables).

\subsection*{Measuring CSAM hosted through the Tor network}

In our study, we crawl webpages hosted on onion services.
According to the Tor Project statistics, there were 693,683 onion domains on 1 January 2023 \citep{tormetrics}.
Onion domains can and do provide any internet service; not all of them host websites.

In practice, we employ parallel crawlers to follow onion links on onion websites, which are subsequently fed to fresh crawlers.
This allows us to continue harvesting in both depth and breadth.
From 2018 to 2023, we collect online content from 176,683 unique onion domain addresses.

We investigate the years 2018--2023 and use a random sample of 10,000 unique onion domains for each year.
Computer-generated random sampling guarantees the genuine randomness of our methods when we refer to random selection in this study.
We then subject the text content to duplicate content filtering, phrase search, and classification.
This returns the detected CSAM percentage for each year, as shown in \autoref{fig:detection}.

To extract only the textual content, we use the \emph{html2text} Python library to convert the HTML pages to plain text representation.
See an example of such a CSAM website in \autoref{fig:detection} and more examples in Supplementary Information A.3.

In our textual representation of websites,
we can see the file names for images and videos,
and also their corresponding caption text
(see detailed examples in Supplementary Methods A.3).
Even websites that offer their full content only after authentication
(see Supplementary Methods A.3, Figure 3)
or behind a paywall serve limited CSAM samples immediately on the landing page.

It would be simple to do true, accurate validation by selecting a random sample of 1,000 distinct domains that --
according to the opening lines of the text -- are unique websites,
and opening these in the Tor Browser to see if they share CSAM\@.
Although this is one possible assessment method and yields the ground truth,
we do not download, open, or view any media content in this research, rather solely focus on textual data.
Accessing CSAM websites would raise ethical, safety, and legal concerns, even in the context of academic research.

Web crawling as a method is biased towards websites that are frequently linked,
and it cannot locate onion websites if there are no links to them.
As a separate issue, the sampling includes onion websites that employ multiple alternative onion addresses.
Despite the fact that we eliminate duplicates,
a website with multiple publicly linked onion domains on other onion websites has not only a greater chance of being crawled but --
through random sampling -- there is also a higher likelihood of being selected for measurements.
Therefore, our methodology favours and estimates popular linked onion websites with several domains
-- not all possible onion websites.
By using a large dataset and continual onion link discovery, we minimise this bias.

\subsubsection*{Manual investigation}
We randomly select the plain text representations of 1,000 onion websites for each year, 2018--2023.
We read the text content of these websites to determine whether they share CSAM and what the English vocabulary
is for this type of page.
Websites that share CSAM make this fact abundantly evident on the front page (see Supplementary Methods A.3),
as well as through the use of explicit, distinct wording.

We have identified \rounding{22.10}\% (N = 221 of 1,000) domain addresses that share CSAM in January 2023.
We repeated this test for the years 2018--2022 (see Supplementary Methods A.3.1) and find that \rounding{19.50}\% (N = 195 of 1,000) of onion domains possess CSAM in 2022,
\rounding{27.20}\% (N = 272 of 1,000) in 2021, and \rounding{19.00}\% (190 of 1,000) in 2020.
While in 2019, \rounding{10.800}\% (N = 108 of 1,000) and in 2018, \rounding{9.000}\% (N = 90 of 1,000) of onion domains shared CSAM\@.

\subsubsection*{Basic keyword search}
Now, we randomly select 10,000 onion websites that were online in December 2022
and perform a basic case-insensitive keyword search (see Supplementary Methods A.1).
This modest matching with 11 explicit CSAM phrases -- including `child porn', `childxxx', `lolita', `preteen', and similar -- produces 2,642 domains from 10,000 onion domains.

When a website has multiple alternate onion domains, we eliminate duplicates.
We execute the search against the content of these distinct websites.
As expected, the algorithm returns a smaller subset -- 2,142 domains that present unique websites.
The search returns 306 matches from these 2,142 domains.

Manually reading the websites, we estimate the false positive (20 from 306, \rounding{6.53594771242}\%) and false negative (\rounding{6.00}\%) rates (see Supplementary Equations D).
According to this keyword-based basic search with the stated false positive and false negative estimation,
\rounding{18.49}\% of unique websites hosted through the Tor network share CSAM\@ in December 2022.

\subsubsection*{Text-based CSAM detection classifiers}

Using the \emph{NLTK} Python library,
we construct a naive Bayes classifier and a decision tree classifier (see Supplementary Methods A.2).
To train classifiers, we manually produce representative CSAM (positive) and other (negative) website datasets.
We curate 1,006 pages from 306 unique CSAM websites and 6,271 pages from 733 unique non-CSAM websites.
These methods have simplistic designs and apparently unrealistic assumptions,
but are known to be accurate for text classification;
a naive Bayes classifier even outperforms sophisticated support vector machines (SVMs)
in text classification, or reaches similar accuracy \citep{DBLP:journals/lalc/Yu08,DBLP:journals/bjmc/PranckeviciusM17}.
For us, they offer a clear benefit: we can understand and interpret them, and after training,
output the detection phrases, combine them, and fine-tune a powerful detection algorithm.

\subsubsection*{Shareable text-based detection for CSAM}

Our goal is to provide search engines with shareable matching phrases so that they can filter CSAM and we can continue to update the phrases.
The classifiers use obvious phrases without much extra logic to match CSAM websites.
This enables us to create a detection algorithm with 404 accurate English phrases (`childxxx', `childrenxxx', `underage slut', etc.).
This is effective, as the vast majority of onion websites are written in English, and search data indicates that almost all users seek explicit material using English terminology.

While selecting the 404 phrases, we only include those that explicitly refer to sexual activity with children;
therefore, we exclude phrases such as `baby love'
-- although it does not generate false positives in the context of Tor.
The inclusion of implicit terminology would give rise to ethical concerns about censorship.

A total of 32.5\% (N = 35,751,619) of searches on \AHMIA{} include sexual phrases, and many of them implicitly might seek CSAM (i.e., `young teen girls sex').
Thus, the creator of the search engine -- the first author of this paper -- decided after the research in November 2023 to filter all sexual and suspicious searches, despite the collateral damage.
This is in response to the widespread search and distribution of illegal child sexual abuse content via Tor, as opposed to legal pornography.

\subsection*{Measuring CSAM searches on the Tor search engine}

We analyse search queries from a well-known public search engine for onion websites.
\AHMIA{} provided us with a list of all search queries from February 2018 to February 2023.
During these five years, search engine users performed 238,794,231 queries.
We analyse these search phrases to determine what Tor users are seeking primarily from onion services.

We conducted limited initial experiments using small-scale interference techniques
with our partner Tor search engine to prevent users from accessing CSAM\@.
Hence, a priori we expected that users would seek little to no CSAM content because
the search engine removes detected CSAM from search results;
redirects users who search for such material using obvious terminology to seek assistance;
and bans any sex-related queries, including legal ones.
Nevertheless, in January 2023, 25 of the top 100 queries seek CSAM content,
despite these previous interference techniques.

We examine searches (N = 238,794,231, \AHMIA{}, February 2018 -- February 2023)
from users seeking content from the Tor network
and discover that explicit CSAM-related search phrases account for \rounding{6.71081413185}\%
of the queries (see separate analysis of individual queries in Supplementary Methods A.7).

\subsection*{Investigating the user's search sessions}

\noindent %
\begin{minipage}[c]{0.35\textwidth}
  We track queries per user.
  We examine the entire search history to follow a total of 110,133,715 search sessions
  and study how many search sessions include at least one search phrase exclusive to underage content (see Supplementary Discussion C.1).

  Even without cookies or IP addresses, it is simple to track a user's searches by looking at the HTTP referring metadata.
  This means that the HTTP request for the new search includes the previous search.
  We assume that a user inputs new searches within five minutes of the last search.
  See an example snippet from the web server logs in \autoref{fig:searchpath}.
\end{minipage}%
\hfill %
\begin{minipage}[c]{0.6\textwidth}
  \includegraphics[width=\linewidth]{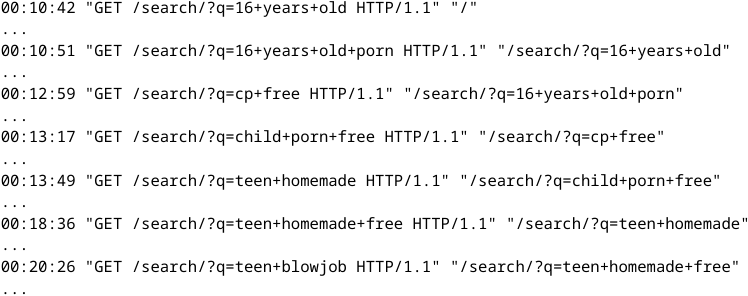}
  \captionof{figure}{User-entered search phrases produce a search session in the HTTP logs.}\label{fig:searchpath}
\end{minipage}

\vspace{5pt}

During this illustrative search session, the user entered seven distinct queries,
some of which reveal that the user is not only interested in
teen sex by adults (age eighteen or nineteen)
but also explicit underage sexual material:
\begin{small}
\begin{displaymath}
\textbf{16 years old}\rightarrow\textbf{16 years old porn}\rightarrow\textbf{cp free}\rightarrow\textbf{child porn free}
\rightarrow\textbf{teen homemade}\rightarrow\textbf{teen homemade free}\rightarrow\textbf{teen blowjob}
\end{displaymath}
\end{small}

\section*{Data availability}\label{sec:ethics}
We believe that scientists should share research data whenever possible (e.g., when it does not violate GDPR or privacy, or when it does not weaponise attacks or sexual abuse).
Our study employs the following three datasets:
(i) website text data from the Tor network,
(ii) web server logs from a search engine,
and (iii) questionnaire responses from CSAM users.
We do not share information that gives direct access to CSAM\@, such as the website text of onion websites (i) that share or link to CSAM\@.
Although the web search logs (ii) do not necessarily contain personal information, it seems feasible to combine or extract such information from the data; thus, we do not release the web search logs.
We release (iii) questionnaire responses from CSAM users and a step-by-step tutorial to help researchers become familiar with our survey data
and replicate our findings (\emph{Zenodo} Open Science repository \url{https://zenodo.org/doi/10.5281/zenodo.10457587}).
Supplementary Information includes Python code for the search using CSAM-related vocabulary,
a Naive Bayes classifier,
and p-value tests to determine the similarity between the distributions.

\section*{Acknowledgments}\label{sec:acknowledgments}

This project has received funding from the European Research Council (ERC) under
the European Union's Horizon 2020 research and innovation programme
under project numbers 804476 (SCARE) and 952622 (SPIRS). %
The survey data was collected under Suojellaan Lapsia, Protect Children ry.
The Safe Online Initiative at End Violence is funding their ReDirection project.

\section*{Author contributions}\label{sec:contributions}

\noindent %
\begin{minipage}[c]{0.36\textwidth}
  J.N. led the article writing and administration.\\
  J.N. operates the \AHMIA{} Tor search engine.\\
  J.N. collected and analysed the Tor website and search engine data.\\
  J.N., N.V.V., and T.I. designed the research.\\
  T.I., A.O., and N.V.V. designed the survey.\\
  M.A. analysed the survey data.\\
  A.P. validated and verified our work and contributed to the figures.\\
  D.A. and J.N. wrote the discussion section.\\
  T.I., J.N., M.A., and B.B. proofread the article.\\
  B.B. acquired funding and edited the article.
\end{minipage}%
\hfill %
\begin{minipage}[c]{0.6\textwidth}
  Contributor Roles Taxonomy (\href{https://credit.niso.org/}{CRediT}) statement of authorship contribution.
  \resizebox{1.0\linewidth}{!}{%
      \begin{tabular}{p{0.35\linewidth}ccccccccccc}
        \toprule
        \textbf{Contributions} &  & \textbf{J.N.} & \textbf{A.P.} & \textbf{B.B.} & \textbf{T.I.} & \textbf{A.O.} & \textbf{V.S.} & \textbf{N.V.V.} & \textbf{M.A.}  & \textbf{D.A.} \ \\
        \midrule
  Conceptualisation & & $\bullet$ & $\bullet$ & $\bullet$ & $\bullet$ & $\bullet$ & $\circ$ & $\bullet$ & $\circ$ & $\circ$ \\
  Data curation & & $\bullet$ & $\circ$ & $\bullet$ & $\circ$ & $\circ$ & $\circ$ & $\circ$ & $\bullet$ & $\circ$ \\
  Formal Analysis & & $\bullet$ & $\bullet$ & $\circ$ & $\circ$ & $\circ$ & $\circ$ & $\circ$ & $\bullet$ & $\circ$ \\
  Funding acquisition & & $\circ$ & $\circ$ & $\bullet$ & $\circ$ & $\circ$ & $\circ$ & $\bullet$ & $\circ$ & $\bullet$ \\
  Investigation & & $\bullet$ & $\circ$ & $\circ$ & $\bullet$ & $\bullet$ & $\bullet$ & $\bullet$ & $\circ$ & $\circ$ \\
  Methodology & & $\bullet$ & $\circ$ & $\circ$ & $\bullet$ & $\bullet$ & $\circ$ & $\bullet$ & $\bullet$ & $\circ$ \\
  Project administration & & $\bullet$ & $\circ$ & $\bullet$ & $\bullet$ & $\circ$ & $\circ$ & $\bullet$ & $\circ$ & $\circ$ \\
  Resources & & $\bullet$ & $\circ$ & $\bullet$ & $\circ$ & $\circ$ & $\circ$ & $\bullet$ & $\circ$ & $\circ$ \\
  Software & & $\bullet$ & $\circ$ & $\circ$ & $\circ$ & $\circ$ & $\circ$ & $\circ$ & $\circ$ & $\circ$ \\
  Supervision & & $\bullet$ & $\circ$ & $\bullet$ & $\circ$ & $\bullet$ & $\circ$ & $\bullet$ & $\circ$ & $\circ$ \\
  Validation & & $\circ$ & $\bullet$ & $\circ$ & $\circ$ & $\circ$ & $\circ$ & $\circ$ & $\bullet$ & $\circ$ \\
  Visualisation & & $\bullet$ & $\bullet$ & $\circ$ & $\circ$ & $\circ$ & $\circ$ & $\circ$ & $\circ$ & $\circ$ \\
  Writing -- original draft & & $\bullet$ & $\bullet$ & $\circ$ & $\bullet$ & $\circ$ & $\bullet$ & $\circ$ & $\bullet$ & $\bullet$ \\
  Writing -- review \& editing & & $\bullet$ & $\bullet$ & $\bullet$ & $\bullet$ & $\circ$ & $\circ$ & $\circ$ & $\bullet$ & $\bullet$ \\
        \bottomrule
        \end{tabular}%
  }
\end{minipage}

\section*{Competing interests}

The authors declare no competing interests.

\section*{Additional information}

See \textbf{Supplementary Information}.


\begin{thebibliography}{10}
\urlstyle{rm}
\expandafter\ifx\csname url\endcsname\relax
  \def\url#1{\texttt{#1}}\fi
\expandafter\ifx\csname urlprefix\endcsname\relax\def\urlprefix{URL }\fi
\expandafter\ifx\csname doiprefix\endcsname\relax\def\doiprefix{DOI: }\fi
\providecommand{\bibinfo}[2]{#2}
\providecommand{\eprint}[2][]{\url{#2}}

\bibitem{DBLP:journals/chb/Guitton13a}
\bibinfo{author}{Guitton, C.}
\newblock \bibinfo{journal}{\bibinfo{title}{{A review of the available content
  on Tor hidden services: The case against further development}}}.
\newblock {\emph{\JournalTitle{Computers in Human Behavior}}}
  \textbf{\bibinfo{volume}{29}}, \bibinfo{pages}{2805--2815},
  \doiprefix\url{https://doi.org/10.1016/j.chb.2013.07.031}
  (\bibinfo{year}{2013}).

\bibitem{DBLP:journals/chb/FoxCL15}
\bibinfo{author}{Fox, J.}, \bibinfo{author}{Cruz, C.} \& \bibinfo{author}{Lee,
  J.~Y.}
\newblock \bibinfo{journal}{\bibinfo{title}{{Perpetuating online sexism
  offline: Anonymity, interactivity, and the effects of sexist hashtags on
  social media}}}.
\newblock {\emph{\JournalTitle{Computers in Human Behavior}}}
  \textbf{\bibinfo{volume}{52}}, \bibinfo{pages}{436--442},
  \doiprefix\url{https://doi.org/10.1016/j.chb.2015.06.024}
  (\bibinfo{year}{2015}).

\bibitem{quayle2004child}
\bibinfo{author}{Quayle, E.} \& \bibinfo{author}{Taylor, M.}
\newblock \emph{\bibinfo{title}{{Child pornography: An internet crime}}}
  (\bibinfo{publisher}{Routledge}, \bibinfo{year}{2004}).

\bibitem{babchishin2015online}
\bibinfo{author}{Babchishin, K.~M.}, \bibinfo{author}{Hanson, R.~K.} \&
  \bibinfo{author}{VanZuylen, H.}
\newblock \bibinfo{journal}{\bibinfo{title}{{Online child pornography offenders
  are different: A meta-analysis of the characteristics of online and offline
  sex offenders against children}}}.
\newblock {\emph{\JournalTitle{Archives of sexual behavior}}}
  \textbf{\bibinfo{volume}{44}}, \bibinfo{pages}{45--66}
  (\bibinfo{year}{2015}).

\bibitem{dsm5}
\bibinfo{title}{{Diagnostic and Statistical Manual of Mental Disorders, Fifth
  Edition, Text Revision (DSM-5-TR)}} (\bibinfo{year}{2022}).
\newblock
  \bibinfo{note}{\url{https://doi.org/10.1176/appi.books.9780890425787}}.

\bibitem{knack2020motivational}
\bibinfo{author}{Knack, N.}, \bibinfo{author}{Holmes, D.} \&
  \bibinfo{author}{Fedoroff, J.~P.}
\newblock \bibinfo{journal}{\bibinfo{title}{{Motivational pathways underlying
  the onset and maintenance of viewing child pornography on the Internet}}}.
\newblock {\emph{\JournalTitle{Behavioral Sciences \& the Law}}}
  \textbf{\bibinfo{volume}{38}}, \bibinfo{pages}{100--116}
  (\bibinfo{year}{2020}).

\bibitem{sharpe2021problematic}
\bibinfo{author}{Sharpe, M.} \& \bibinfo{author}{Mead, D.}
\newblock \bibinfo{journal}{\bibinfo{title}{Problematic pornography use: Legal
  and health policy considerations}}.
\newblock {\emph{\JournalTitle{Current Addiction Reports}}}
  \bibinfo{pages}{1--12} (\bibinfo{year}{2021}).

\bibitem{insoll2022}
\bibinfo{author}{Insoll, T.}, \bibinfo{author}{Ovaska, A.~K.},
  \bibinfo{author}{Nurmi, J.}, \bibinfo{author}{Aaltonen, M.} \&
  \bibinfo{author}{Vaaranen-Valkonen, N.}
\newblock \bibinfo{journal}{\bibinfo{title}{{Risk Factors for Child Sexual
  Abuse Material Users Contacting Children Online: Results of an Anonymous
  Multilingual Survey on the Dark Web}}}.
\newblock {\emph{\JournalTitle{Journal of Online Trust and Safety}}}
  \textbf{\bibinfo{volume}{1}} (\bibinfo{year}{2022}).

\bibitem{levine2022increasing}
\bibinfo{author}{Levine, B.~N.}
\newblock \bibinfo{title}{{Increasing the Efficacy of Investigations of Online
  Child Sexual Exploitation}} (\bibinfo{year}{2022}).
\newblock
  \bibinfo{note}{\url{https://www.ojp.gov/library/publications/increasing-efficacy-investigations-online-child-sexual-exploitation-report}}.

\bibitem{DBLP:conf/soups/GallagherPM17}
\bibinfo{author}{Gallagher, K.}, \bibinfo{author}{Patil, S.} \&
  \bibinfo{author}{Memon, N.~D.}
\newblock \bibinfo{title}{{New Me: Understanding Expert and Non-Expert
  Perceptions and Usage of the Tor Anonymity Network}}.
\newblock In \emph{\bibinfo{booktitle}{Thirteenth Symposium on Usable Privacy
  and Security, {SOUPS} 2017, Santa Clara, CA, USA, July 12-14, 2017}},
  \bibinfo{pages}{385--398} (\bibinfo{publisher}{{USENIX} Association},
  \bibinfo{year}{2017}).

\bibitem{DBLP:conf/soups/Obada-ObiehSB20}
\bibinfo{author}{Obada{-}Obieh, B.}, \bibinfo{author}{Spagnolo, L.} \&
  \bibinfo{author}{Beznosov, K.}
\newblock \bibinfo{title}{{Towards Understanding Privacy and Trust in Online
  Reporting of Sexual Assault}}.
\newblock In \bibinfo{editor}{Lipford, H.~R.} \& \bibinfo{editor}{Chiasson, S.}
  (eds.) \emph{\bibinfo{booktitle}{Sixteenth Symposium on Usable Privacy and
  Security, {SOUPS} 2020, August 7-11, 2020}}, \bibinfo{pages}{145--164}
  (\bibinfo{publisher}{{USENIX} Association}, \bibinfo{year}{2020}).

\bibitem{DBLP:conf/uss/WinterERDCF18}
\bibinfo{author}{Winter, P.} \emph{et~al.}
\newblock \bibinfo{title}{{How Do Tor Users Interact With Onion Services?}}
\newblock In \bibinfo{editor}{Enck, W.} \& \bibinfo{editor}{Felt, A.~P.} (eds.)
  \emph{\bibinfo{booktitle}{27th {USENIX} Security Symposium, {USENIX} Security
  2018, Baltimore, MD, USA, August 15-17, 2018}}, \bibinfo{pages}{411--428}
  (\bibinfo{publisher}{{USENIX} Association}, \bibinfo{year}{2018}).

\bibitem{DBLP:conf/websci/ZabihimayvanSDA19}
\bibinfo{author}{Zabihimayvan, M.}, \bibinfo{author}{Sadeghi, R.},
  \bibinfo{author}{Doran, D.} \& \bibinfo{author}{Allahyari, M.}
\newblock \bibinfo{title}{{A Broad Evaluation of the Tor English Content
  Ecosystem}}.
\newblock In \bibinfo{editor}{Boldi, P.} \emph{et~al.} (eds.)
  \emph{\bibinfo{booktitle}{Proceedings of the 11th {ACM} Conference on Web
  Science, WebSci 2019, Boston, MA, USA, June 30 - July 03, 2019}},
  \bibinfo{pages}{333--342},
  \doiprefix\url{https://doi.org/10.1145/3292522.3326031}
  (\bibinfo{publisher}{{ACM}}, \bibinfo{year}{2019}).

\bibitem{doi:10.1073/pnas.2011893117}
\bibinfo{author}{Jardine, E.}, \bibinfo{author}{Lindner, A.~M.} \&
  \bibinfo{author}{Owenson, G.}
\newblock \bibinfo{journal}{\bibinfo{title}{{The potential harms of the Tor
  anonymity network cluster disproportionately in free countries}}}.
\newblock {\emph{\JournalTitle{Proceedings of the National Academy of
  Sciences}}} \textbf{\bibinfo{volume}{117}}, \bibinfo{pages}{31716--31721},
  \doiprefix\url{https://doi.org/10.1073/pnas.2011893117}
  (\bibinfo{year}{2020}).

\bibitem{aked2011investigation}
\bibinfo{author}{Aked, S.}
\newblock \bibinfo{journal}{\bibinfo{title}{{An investigation into darknets and
  the content available via anonymous peer-to-peer file sharing}}}.
\newblock {\emph{\JournalTitle{9th Australian Information Security Management
  Conference}}}  (\bibinfo{year}{2011}).

\bibitem{DBLP:journals/istr/Figueras-Martin22}
\bibinfo{author}{Figueras{-}Mart{\'{\i}}n, E.},
  \bibinfo{author}{Mag{\'{a}}n{-}Carri{\'{o}}n, R.} \&
  \bibinfo{author}{Boubeta{-}Puig, J.}
\newblock \bibinfo{journal}{\bibinfo{title}{{Drawing the web structure and
  content analysis beyond the Tor darknet: Freenet as a case of study}}}.
\newblock {\emph{\JournalTitle{J. Inf. Secur. Appl.}}}
  \textbf{\bibinfo{volume}{68}}, \bibinfo{pages}{103229},
  \doiprefix\url{https://doi.org/10.1016/j.jisa.2022.103229}
  (\bibinfo{year}{2022}).

\bibitem{DBLP:journals/iet-ifs/OwenS16}
\bibinfo{author}{Owen, G.} \& \bibinfo{author}{Savage, N.}
\newblock \bibinfo{journal}{\bibinfo{title}{{Empirical analysis of Tor Hidden
  Services}}}.
\newblock {\emph{\JournalTitle{{IET} Information Security}}}
  \textbf{\bibinfo{volume}{10}}, \bibinfo{pages}{113--118},
  \doiprefix\url{https://doi.org/10.1049/iet-ifs.2015.0121}
  (\bibinfo{year}{2016}).

\bibitem{DBLP:conf/isi/SpittersVS14}
\bibinfo{author}{Spitters, M.}, \bibinfo{author}{Verbruggen, S.} \&
  \bibinfo{author}{van Staalduinen, M.}
\newblock \bibinfo{title}{{Towards a Comprehensive Insight into the Thematic
  Organization of the Tor Hidden Services}}.
\newblock In \emph{\bibinfo{booktitle}{{IEEE} Joint Intelligence and Security
  Informatics Conference, {JISIC} 2014, The Hague, The Netherlands, 24-26
  September, 2014}}, \bibinfo{pages}{220--223},
  \doiprefix\url{https://doi.org/10.1109/JISIC.2014.40}
  (\bibinfo{publisher}{{IEEE}}, \bibinfo{year}{2014}).

\bibitem{DBLP:journals/di/DalinsWC18}
\bibinfo{author}{Dalins, J.}, \bibinfo{author}{Wilson, C.} \&
  \bibinfo{author}{Carman, M.~J.}
\newblock \bibinfo{journal}{\bibinfo{title}{Criminal motivation on the dark
  web: {A} categorisation model for law enforcement}}.
\newblock {\emph{\JournalTitle{Digit. Investig.}}}
  \textbf{\bibinfo{volume}{24}}, \bibinfo{pages}{62--71},
  \doiprefix\url{https://doi.org/10.1016/j.diin.2017.12.003}
  (\bibinfo{year}{2018}).

\bibitem{he2019classification}
\bibinfo{author}{He, S.}, \bibinfo{author}{He, Y.} \& \bibinfo{author}{Li, M.}
\newblock \bibinfo{title}{Classification of illegal activities on the dark
  web}.
\newblock In \emph{\bibinfo{booktitle}{Proceedings of the 2nd International
  Conference on Information Science and Systems}}, \bibinfo{pages}{73--78}
  (\bibinfo{year}{2019}).

\bibitem{harris2021child}
\bibinfo{author}{Harris, M.}, \bibinfo{author}{Allardyce, S.} \&
  \bibinfo{author}{Findlater, D.}
\newblock \bibinfo{journal}{\bibinfo{title}{Child sexual abuse and covid-19:
  Side effects of changed societies and positive lessons for prevention}}.
\newblock {\emph{\JournalTitle{Criminal Behaviour and Mental Health}}}
  \textbf{\bibinfo{volume}{31}}, \bibinfo{pages}{289} (\bibinfo{year}{2021}).

\bibitem{steel2021collecting}
\bibinfo{author}{Steel, C.~M.}, \bibinfo{author}{Newman, E.},
  \bibinfo{author}{O'Rourke, S.} \& \bibinfo{author}{Quayle, E.}
\newblock \bibinfo{journal}{\bibinfo{title}{Collecting and viewing behaviors of
  child sexual exploitation material offenders}}.
\newblock {\emph{\JournalTitle{Child Abuse \& Neglect}}}
  \textbf{\bibinfo{volume}{118}}, \bibinfo{pages}{105133}
  (\bibinfo{year}{2021}).

\bibitem{blanchard2009pedophilia}
\bibinfo{author}{Blanchard, R.} \emph{et~al.}
\newblock \bibinfo{journal}{\bibinfo{title}{Pedophilia, hebephilia, and the
  {DSM-V}}}.
\newblock {\emph{\JournalTitle{Archives of sexual behavior}}}
  \textbf{\bibinfo{volume}{38}}, \bibinfo{pages}{335--350}
  (\bibinfo{year}{2009}).

\bibitem{anderson2003relative}
\bibinfo{author}{Anderson, S.~E.}, \bibinfo{author}{Dallal, G.~E.} \&
  \bibinfo{author}{Must, A.}
\newblock \bibinfo{journal}{\bibinfo{title}{Relative weight and race influence
  average age at menarche: results from two nationally representative surveys
  of us girls studied 25 years apart}}.
\newblock {\emph{\JournalTitle{Pediatrics}}} \textbf{\bibinfo{volume}{111}},
  \bibinfo{pages}{844--850} (\bibinfo{year}{2003}).

\bibitem{lin2020exposure}
\bibinfo{author}{Lin, W.-H.}, \bibinfo{author}{Liu, C.-H.} \&
  \bibinfo{author}{Yi, C.-C.}
\newblock \bibinfo{journal}{\bibinfo{title}{Exposure to sexually explicit media
  in early adolescence is related to risky sexual behavior in emerging
  adulthood}}.
\newblock {\emph{\JournalTitle{PloS one}}} \textbf{\bibinfo{volume}{15}},
  \bibinfo{pages}{e0230242} (\bibinfo{year}{2020}).

\bibitem{brown2009x}
\bibinfo{author}{Brown, J.~D.} \& \bibinfo{author}{L'Engle, K.~L.}
\newblock \bibinfo{journal}{\bibinfo{title}{{X-rated: Sexual attitudes and
  behaviors associated with US early adolescents' exposure to sexually explicit
  media}}}.
\newblock {\emph{\JournalTitle{Communication research}}}
  \textbf{\bibinfo{volume}{36}}, \bibinfo{pages}{129--151}
  (\bibinfo{year}{2009}).

\bibitem{hunter2010developmental}
\bibinfo{author}{Hunter, J.~A.}, \bibinfo{author}{Figueredo, A.~J.} \&
  \bibinfo{author}{Malamuth, N.~M.}
\newblock \bibinfo{journal}{\bibinfo{title}{Developmental pathways into social
  and sexual deviance}}.
\newblock {\emph{\JournalTitle{Journal of Family Violence}}}
  \textbf{\bibinfo{volume}{25}}, \bibinfo{pages}{141--148}
  (\bibinfo{year}{2010}).

\bibitem{asmussen2020adverse}
\bibinfo{author}{Asmussen, K.}, \bibinfo{author}{Fischer, F.},
  \bibinfo{author}{Drayton, E.} \& \bibinfo{author}{McBride, T.}
\newblock \bibinfo{journal}{\bibinfo{title}{{Adverse childhood experiences:
  What we know, what we don’t know, and what should happen next}}}.
\newblock {\emph{\JournalTitle{Early intervention foundation}}}
  \textbf{\bibinfo{volume}{129}} (\bibinfo{year}{2020}).

\bibitem{beier2009can}
\bibinfo{author}{Beier, K.~M.} \emph{et~al.}
\newblock \bibinfo{journal}{\bibinfo{title}{Can pedophiles be reached for
  primary prevention of child sexual abuse? first results of the berlin
  prevention project dunkelfeld (ppd)}}.
\newblock {\emph{\JournalTitle{The journal of forensic psychiatry \&
  psychology}}} \textbf{\bibinfo{volume}{20}}, \bibinfo{pages}{851--867}
  (\bibinfo{year}{2009}).

\bibitem{icd11}
\bibinfo{title}{{The International Classification of Diseases 11th Revision
  (ICD-11) -- 6C72 Compulsive sexual behaviour disorder}}
  (\bibinfo{year}{2022}).
\newblock
  \bibinfo{note}{\url{https://icd.who.int/browse11/l-m/en\#/http://id.who.int/icd/entity/1630268048}}.

\bibitem{laier2013cybersex}
\bibinfo{author}{Laier, C.}, \bibinfo{author}{Pawlikowski, M.},
  \bibinfo{author}{Pekal, J.}, \bibinfo{author}{Schulte, F.~P.} \&
  \bibinfo{author}{Brand, M.}
\newblock \bibinfo{journal}{\bibinfo{title}{{Cybersex addiction: Experienced
  sexual arousal when watching pornography and not real-life sexual contacts
  makes the difference}}}.
\newblock {\emph{\JournalTitle{Journal of behavioral addictions}}}
  \textbf{\bibinfo{volume}{2}}, \bibinfo{pages}{100--107}
  (\bibinfo{year}{2013}).

\bibitem{quayle2003model}
\bibinfo{author}{Quayle, E.} \& \bibinfo{author}{Taylor, M.}
\newblock \bibinfo{journal}{\bibinfo{title}{{Model of problematic Internet use
  in people with a sexual interest in children}}}.
\newblock {\emph{\JournalTitle{CyberPsychology \& Behavior}}}
  \textbf{\bibinfo{volume}{6}}, \bibinfo{pages}{93--106}
  (\bibinfo{year}{2003}).

\bibitem{quayle2006sex}
\bibinfo{author}{Quayle, E.}, \bibinfo{author}{Vaughan, M.} \&
  \bibinfo{author}{Taylor, M.}
\newblock \bibinfo{journal}{\bibinfo{title}{{Sex offenders, Internet child
  abuse images and emotional avoidance: The importance of values}}}.
\newblock {\emph{\JournalTitle{Aggression and violent Behavior}}}
  \textbf{\bibinfo{volume}{11}}, \bibinfo{pages}{1--11} (\bibinfo{year}{2006}).

\bibitem{levenson2017obstacles}
\bibinfo{author}{Levenson, J.~S.}, \bibinfo{author}{Willis, G.~M.} \&
  \bibinfo{author}{Vicencio, C.~P.}
\newblock \bibinfo{journal}{\bibinfo{title}{{Obstacles to help-seeking for
  sexual offenders: Implications for prevention of sexual abuse}}}.
\newblock {\emph{\JournalTitle{Journal of child sexual abuse}}}
  \textbf{\bibinfo{volume}{26}}, \bibinfo{pages}{99--120}
  (\bibinfo{year}{2017}).

\bibitem{schmidt2022outpatient}
\bibinfo{author}{Schmidt, A.~F.} \& \bibinfo{author}{Niehaus, S.}
\newblock \bibinfo{journal}{\bibinfo{title}{{Outpatient therapists’
  perspectives on working with persons who are sexually interested in
  minors}}}.
\newblock {\emph{\JournalTitle{Archives of sexual behavior}}}
  \textbf{\bibinfo{volume}{51}}, \bibinfo{pages}{4157--4178}
  (\bibinfo{year}{2022}).

\bibitem{grady2019can}
\bibinfo{author}{Grady, M.~D.}, \bibinfo{author}{Levenson, J.~S.},
  \bibinfo{author}{Mesias, G.}, \bibinfo{author}{Kavanagh, S.} \&
  \bibinfo{author}{Charles, J.}
\newblock \bibinfo{journal}{\bibinfo{title}{{``I can’t talk about that'':
  Stigma and fear as barriers to preventive services for minor-attracted
  persons.}}}
\newblock {\emph{\JournalTitle{Stigma and Health}}}
  \textbf{\bibinfo{volume}{4}}, \bibinfo{pages}{400} (\bibinfo{year}{2019}).

\bibitem{letourneau2014need}
\bibinfo{author}{Letourneau, E.~J.}, \bibinfo{author}{Eaton, W.~W.},
  \bibinfo{author}{Bass, J.}, \bibinfo{author}{Berlin, F.~S.} \&
  \bibinfo{author}{Moore, S.~G.}
\newblock \bibinfo{title}{The need for a comprehensive public health approach
  to preventing child sexual abuse} (\bibinfo{year}{2014}).

\bibitem{nebhinani2012reasons}
\bibinfo{author}{Nebhinani, N.}, \bibinfo{author}{Sarkar, S.},
  \bibinfo{author}{Ghai, S.} \& \bibinfo{author}{Basu, D.}
\newblock \bibinfo{journal}{\bibinfo{title}{{Reasons for help-seeking and
  associated fears in subjects with substance dependence}}}.
\newblock {\emph{\JournalTitle{Indian journal of psychological medicine}}}
  \textbf{\bibinfo{volume}{34}}, \bibinfo{pages}{153--158}
  (\bibinfo{year}{2012}).

\bibitem{collier22}
\bibinfo{author}{Collier, B.} \& \bibinfo{author}{Stewart, J.}
\newblock \bibinfo{journal}{\bibinfo{title}{{Privacy Worlds: Exploring Values
  and Design in the Development of the Tor Anonymity Network}}}.
\newblock {\emph{\JournalTitle{Science, Technology, \& Human Values}}}
  \textbf{\bibinfo{volume}{47}}, \bibinfo{pages}{910--936},
  \doiprefix\url{https://doi.org/10.1177/01622439211039019}
  (\bibinfo{year}{2022}).

\bibitem{sarda2023onion}
\bibinfo{author}{Sard{\'a}, T.}
\newblock \bibinfo{journal}{\bibinfo{title}{{An onion with layers of hope and
  fear: A cross-case analysis of the media representation of Tor Network
  reflecting theoretical perspectives of new technologies}}}.
\newblock {\emph{\JournalTitle{Security and Privacy}}} \bibinfo{pages}{e296}
  (\bibinfo{year}{2023}).

\bibitem{icme2018}
\bibinfo{author}{{ICMEC}}.
\newblock \bibinfo{title}{{Child Sexual Abuse Material: Model Legislation and
  Global Review. 9th Edition.}} (\bibinfo{year}{2018}).
\newblock
  \bibinfo{note}{\url{https://www.icmec.org/child-pornography-model-legislation-report/}}.

\bibitem{clarke1995situational}
\bibinfo{author}{Clarke, R.~V.}
\newblock \bibinfo{journal}{\bibinfo{title}{Situational crime prevention}}.
\newblock {\emph{\JournalTitle{Crime and justice}}}
  \textbf{\bibinfo{volume}{19}}, \bibinfo{pages}{91--150}
  (\bibinfo{year}{1995}).

\bibitem{HO2022102611}
\bibinfo{author}{Ho, H.}, \bibinfo{author}{Ko, R.} \&
  \bibinfo{author}{Mazerolle, L.}
\newblock \bibinfo{journal}{\bibinfo{title}{{Situational Crime Prevention (SCP)
  techniques to prevent and control cybercrimes: A focused systematic
  review}}}.
\newblock {\emph{\JournalTitle{Computers \& Security}}}
  \textbf{\bibinfo{volume}{115}}, \bibinfo{pages}{102611},
  \doiprefix\url{https://doi.org/10.1016/j.cose.2022.102611}
  (\bibinfo{year}{2022}).

\bibitem{doi:10.1080/13552600.2019.1705925}
\bibinfo{author}{Krone, T.} \emph{et~al.}
\newblock \bibinfo{journal}{\bibinfo{title}{Child sexual abuse material in
  child-centred institutions: situational crime prevention approaches}}.
\newblock {\emph{\JournalTitle{Journal of Sexual Aggression}}}
  \textbf{\bibinfo{volume}{26}}, \bibinfo{pages}{91--110},
  \doiprefix\url{https://doi.org/10.1080/13552600.2019.1705925}
  (\bibinfo{year}{2020}).

\bibitem{prichard2022effects}
\bibinfo{author}{Prichard, J.} \emph{et~al.}
\newblock \bibinfo{journal}{\bibinfo{title}{Effects of automated messages on
  internet users attempting to access ``barely legal'' pornography}}.
\newblock {\emph{\JournalTitle{Sexual Abuse}}} \textbf{\bibinfo{volume}{34}},
  \bibinfo{pages}{106--124} (\bibinfo{year}{2022}).

\bibitem{latth2022effects}
\bibinfo{author}{L{\"a}tth, J.} \emph{et~al.}
\newblock \bibinfo{journal}{\bibinfo{title}{{Effects of internet-delivered
  cognitive behavioral therapy on use of child sexual abuse material: A
  randomized placebo-controlled trial on the Darknet}}}.
\newblock {\emph{\JournalTitle{Internet Interventions}}}
  \textbf{\bibinfo{volume}{30}}, \bibinfo{pages}{100590},
  \doiprefix\url{https://doi.org/10.1016/j.invent.2022.100590}
  (\bibinfo{year}{2022}).

\bibitem{tormetrics}
\bibinfo{title}{{Tor} metrics -- onion services} (\bibinfo{year}{2023}).
\newblock \bibinfo{note}{{The Tor Project}.
  \url{https://metrics.torproject.org/hidserv-dir-v3-onions-seen.html?start=2022-11-23&end=2023-01-03}}.

\bibitem{DBLP:journals/lalc/Yu08}
\bibinfo{author}{Yu, B.}
\newblock \bibinfo{journal}{\bibinfo{title}{{An evaluation of text
  classification methods for literary study}}}.
\newblock {\emph{\JournalTitle{Lit. Linguistic Comput.}}}
  \textbf{\bibinfo{volume}{23}}, \bibinfo{pages}{327--343},
  \doiprefix\url{https://doi.org/10.1093/llc/fqn015} (\bibinfo{year}{2008}).

\bibitem{DBLP:journals/bjmc/PranckeviciusM17}
\bibinfo{author}{Pranckevicius, T.} \& \bibinfo{author}{Marcinkevicius, V.}
\newblock \bibinfo{journal}{\bibinfo{title}{{Comparison of Naive Bayes, Random
  Forest, Decision Tree, Support Vector Machines, and Logistic Regression
  Classifiers for Text Reviews Classification}}}.
\newblock {\emph{\JournalTitle{Balt. J. Mod. Comput.}}}
  \textbf{\bibinfo{volume}{5}},
  \doiprefix\url{https://doi.org/10.22364/bjmc.2017.5.2.05}
  (\bibinfo{year}{2017}).

\end{thebibliography}
\end{document}